\documentclass[useAMS,usenatbib,usegraphicx]{mn2e}

\usepackage{amsmath}
\usepackage{amssymb}
\usepackage[dvips]{color} 

\newcommand{\Msun}{\ensuremath{\mathit{M}_{\odot}}}
\newcommand{\Rsun}{\ensuremath{\mathit{R}_{\odot}}}
\newcommand{\ej}{\mathrm{ej}}
\newcommand{\Ni}{$^{56}$Ni}

\title[Core-collapse SNe brightened by binary companions]
{
Observable fractions of core-collapse supernova light curves 
brightened by binary companions
}
\author[T. J. Moriya, Z.-W. Liu, and R. G. Izzard]
{Takashi J. Moriya$^1$\thanks{moriyatk@astro.uni-bonn.de},
 Zheng-Wei Liu$^1$, and
 Robert G. Izzard$^{1,2}$ \\
$^{1}$
Argelander Institute for Astronomy, University of Bonn,
Auf dem H\"ugel 71, 53121 Bonn, Germany \\
$^{2}$
Institute of Astronomy, University of Cambridge,
Madingley Road, Cambridge, CB3 0HA, United Kingdom
}
\begin{document}

\date{
Accepted, 23 April 2015. Received, 1 April 2015; in original form, 6 February 2015
}

\pagerange{\pageref{firstpage}--\pageref{lastpage}} \pubyear{2015}

\maketitle

\label{firstpage}

\begin{abstract}
Many core-collapse supernova progenitors are presumed to be in binary
systems. If a star explodes in a binary system, the early supernova light curve
can be brightened by the collision of the supernova ejecta with the
companion star.
The early brightening can be observed when the
observer is in the direction of the hole created by the collision.
Based on a population synthesis model, we estimate the
fractions of core-collapse supernovae in which 
the light-curve brightening by the collision can be observed.
We find that $0.19$\% of core-collapse supernova light curves
can be observed with the collisional brightening.
Type~Ibc supernova light curves are more likely to be
brightened by the collision (0.53\%) because of the high fraction of
the progenitors being in binary systems and their proximity to the
companion stars. Type~II and IIb supernova light curves are less
affected ($\sim 10^{-3}$\% and $\sim 10^{-2}$\%, respectively).
Although the early, slow light-curve declines of some 
Type~IIb and Ibc supernovae are argued
to be caused by the collision with the companion star (e.g. SN~2008D),
the small expected fraction, as well as the unrealistically small
separation required, disfavour the argument.
The future transient survey by the Large Synoptic Survey Telescope 
is expected to detect $\sim 10$ Type Ibc supernovae 
with the early collisional brightening per year,
and they will be able to
provide information on supernova progenitors in binary systems.
\end{abstract}


\section{Introduction}\label{introduction}
Core-collapse supernovae (SNe) are explosions of massive stars.
Massive stars do not explode as they are born. 
During their evolution to their death, which takes millions of years,
the internal structures of the
stars dramatically change and their masses can be reduced because of
mass loss. They can also be affected by their binary companions.
\citet{sana2012} found that more than 70\% of massive stars
which can eventually explode are once in close binary systems.
The effect of stellar duplicity on the stellar evolution has been
studied (see \citealt{langer2012} for a review).
Especially, core-collapse SNe with little or no hydrogen
(SNe~IIL/IIb/Ib/Ic) have been suggested to come from binary systems
(e.g.
\citealt{podsiadlowski1992,podsiadlowski1993,maund2004,eldridge2008,yoon2010,folatelli2014,eldridge2014}).
SN light-curve (LC) modelling also supports this idea
(e.g. \citealt{utrobin1994,nomoto1993,woosley1994,blinnikov1998,bersten2012,bersten2014,fremling2014}).

When a SN explosion occurs in a binary system, the SN and the companion
star can be both affected by the explosion
(e.g. \citealt{wheeler1975,fryxell1981,chugai1986,livne1992,marietta2000,pakmor2008,chevalier2012,liu2013,hirai2014,maeda2014,sabach2014,tauris2015}).
Several SN properties can be strongly affected by the existence of the companion.
In particular, \citet[][K10 hereafter]{kasen2010} showed that the binary
companion can alter the early X-ray/ultra-violet/optical SN LCs.
The early optical brightness is increased because of 
the extra heating provided by the collision between the SN ejecta and the companion star.

While early SN optical LCs are brightened by a collision,
they are presumed to decline much more slowly than those
without a collision (K10).
Relative to a normal, e.g. single-star, SN, the LC describing a
collision between SN ejecta and a companion star is brighter (K10).
Energy from the collision counteracts adiabatic cooling after the shock
breakout, thus slowing the LC decline.
The LC of SN~Ib~2008D was related to a collision
because of the extremely slow LC decline during the 5 days after the explosion
\citep{dessart2011}.
SN LC simulations show that
the early LC declines in SNe~Ibc caused by the adiabatic cooling
in the SN ejecta only last about 1 day 
because their progenitors are compact (Wolf-Rayet stars).
The LCs then become flat for several days because of helium recombination
if a sufficient amount of helium is in the progenitors
\citep{dessart2011}.
The declining phase of the LC of SN~2008D lasted for about 5 days and
is difficult to explain with standard compact Wolf-Rayet progenitors.
Because many SN~Ibc progenitors are suggested to be in binary systems,
\citet{dessart2011} proposed that the slow
decline can be related to the collision of the SN ejecta with the binary companion.
However, \citet{bersten2013} argued that
the separation required to explain the early luminosity
of SN~2008D is unrealistically small ($\sim \Rsun$) 
and it is not likely caused by the collision.
They suggested that the \Ni\ mixing to the outermost layers of
the SN ejecta lead to the early, slow LC decline rather than the
collision.
Other plausible explanations for the slow LC decline include
a large progenitor radius (e.g. \citealt{chevalier2008,nakar2014})
and the existence of an extended dense circumstellar medium
(e.g. \citealt{balberg2011,svirski2014}).

\begin{figure}
\centering
\includegraphics[width=0.5\columnwidth]{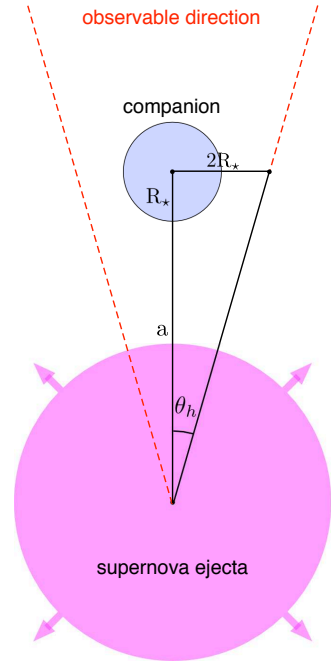}
\caption{
Schematic picture of important parameters in our model.
When the SN ejecta collides with the companion star with radius $R_\star$
at a separation $a$, a hole with the half opening angle of
$\theta_h\simeq \arctan (2R_\star/a)$ is created by the collision. 
The early LC brightening because of the collision
can be observed when the observer is in the direction of the hole (K10).
}
\label{collision}
\end{figure}

While a large fraction of core-collapse SN progenitors may be affected
by the binary companion during their evolution,
whether we can observe brightening in the early LCs is not obvious
even if the progenitors are in binary systems.
This is because not only the progenitors need to be in binary systems
but also observers need to be within certain limited viewing angles
from which the brightening can be observed (K10, Fig.~\ref{collision}).
In this paper, we investigate the expected fractions of core-collapse SNe
whose observed LCs are brightened by the collision
based on a population synthesis model.
Our predicted fractions can be tested with future transient surveys.

The rest of this paper is organized as follows.
In Section~\ref{population}, we summarize our population synthesis
model and the general SN properties predicted from the model.
Then, we estimate the fractions of SNe in which the early LC brightening
by the collision is observable in Section~\ref{fraction}.
We discuss our results in Section~\ref{discussion} and conclude
in Section~\ref{conclusion}.

\section{Population synthesis}\label{population}
In this section, we briefly summarize our population synthesis modelling.
We discuss the general SN properties
predicted by our population synthesis model.
We also discuss the uncertainties in our predictions.

\subsection{Method}\label{method}
We model stellar evolution with the binary evolution code
\texttt{binary\_c/nucsyn}
\citep{izzard2004b,izzard2006,izzard2009}
which is a modified and updated version
of that presented by \citet{hurley2002}. We refer to these papers
for the detailed physics and assumptions in the binary evolution model.
Of the key parameters in binary-star evolution, we set the common-envelope
ejection efficiency, $\alpha_{\mathrm{CE}}$, to $1$ by default and
use the fits to detailed models describing the envelope binding energy
parameter, $\lambda$ \citep{dewi2000,tauris2001}.
$\alpha_\mathrm{CE}$ is constrained to be around $\sim 0.1- 1$
for low- and intermediate-mass stars
\citep[e.g.][]{zorotovic2010,demarco2011,davis2012}.
Wind mass loss is taken into account by following the
prescription summarized in \citet{izzard2006}.
A SN is assumed to occur when the carbon-oxygen core mass reaches
a given maximum mass (see \citealt{hurley2000} for details).
When a star explodes in a binary system, the companion star is assumed
to receive a kick velocity from a Maxwellian distribution with a
dispersion of 190 $\mathrm{km~s^{-1}}$, as is assumed in \citet{hurley2002}.

Our stellar populations are modelled by a three-dimensional parameter
space of the primary mass, secondary mass, and orbital period. The
primary mass is distributed according to the initial mass function
of \citet{kroupa2001}, and the secondary mass is chosen to
be flat in the mass ratio $q=M_{2}/M_{1}$. The period distribution
follows that of \citet{sana2012} for O-type
primary stars in excess of $16{\, M_{\odot}}$,
\citet{raghavan2010} for solar-like primary stars with mass below $1.15{\, M_{\odot}}$,
and a fitting function interpolated linearly in primary mass between
these two limiting masses. We evolve systems with primary masses from
$3$~to~$80{\, M_{\odot}}$, mass ratios between $0.1M_\odot$/$M_1$ 
and $1.0$ and orbital periods between $0.1$ and $10^{10}\,\mathrm{days}$.
All our stars have metallicity $Z=0.02$ and start in circular orbits.

The fraction of stars in binary systems in our stellar populations is
based on \citet{raghavan2010}.
All the stars which are O-type or B-type at the zero-age main sequence are
assumed to be in binary systems initially.
Therefore, the initial binary fraction of our SN progenitors is essentially 100\%.
Observations indicate that the fraction of OB stars in binary systems are
more than 70\% \citep{raghavan2010,sana2012}.
Single stars in these observations may be observed as
single because they are merged stars or 
their companion stars have already exploded.
Thus, the actual initial binary fractions are not well-known and
we simply assume the 100\% binary fraction in this study.

\subsection{Supernova properties}\label{snproperties}
The population synthesis model described in the previous section
predicts several SN properties. The total core-collapse SN rate of the Galaxy
predicted from the model is
$(0.93-1.99) \times 10^{-2}\ \mathrm{yr^{-1}}$,
assuming a Galactic star-formation rate of $0.68-1.45$
$M_\odot~\mathrm{yr^{-1}}$ \citep{robitaille2010} and the average stellar mass
from the Kroupa initial mass function (0.83 $M_\odot$).
This rate is consistent with the estimated Galactic core-collapse SN
rate from recent surveys
[e.g. $(2.30\pm 0.48)\times 10^{-2}$ $\mathrm{yr^{-1}}$, \citealt{li2011}].

We determine the SN type of the exploding stars based on the stellar composition.
While there exist many types of core-collapse SNe, we only consider
three SN types (II, IIb, and Ibc) for simplicity.
If hydrogen remains in SN progenitors, the SNe are classified as Type~II
or IIb.
We try to identify Type~IIb because several SN~IIb progenitors are
suggested to be in binary systems
(e.g. \citealt{maund2004,bersten2012,fox2014,folatelli2014}).
The difference between the progenitors of the two SN types is
presumed to be in the remaining hydrogen-rich envelope mass, but the exact mass
dividing the two types is not well-constrained.
In our standard model,
we classify the progenitors with the hydrogen-rich envelope mass
smaller than 4.5 \Msun\ as Type~IIb and the other hydrogen-rich
progenitors as Type~II. 
The dividing mass 4.5 \Msun\ is chosen so that the Type~IIb
fraction in our model is similar to the observed fraction
($\sim 10$\%, \citealt{smith2011,eldridge2013}).
However, the dividing mass 4.5 \Msun\ is higher than
the 2 \Msun\ used in other
progenitor studies (e.g. \citealt{heger2003,groh2013})
or those estimated from observational properties
($\sim 0.5-0.1\ \Msun$, e.g. \citealt{nomoto1993,podsiadlowski1993,woosley1994,blinnikov1998,elmhamdi2006,bersten2012}).
We also show the results with the dividing mass of 2, 0.5, and 0.1 \Msun.
As is shown in the following sections, the SN~IIb fractions
in core-collapse SNe are affected
by the dividing mass but the observational fractions of the early LC
brightening because of the collision are not much.
If a SN is hydrogen-free, it is classified as Type~Ibc.
Because the differences in the progenitors of Type~Ib and Ic SNe may not
simply come from the remaining helium mass and they are not
well-known (e.g. \citealt{dessart2012,hachinger2012}),
we do not sub-divide the hydrogen-free SNe. 

The fractions of each SN type are shown in Table~\ref{table1}.
The fractions of hydrogen-rich SNe (Types~II and IIb) and
the hydrogen-free SNe (Type~Ibc) obtained from our population
synthesis model match the observations reasonably
(Type~II including all the subtypes $\sim 70$\% and Type~Ibc $\sim 30$\%,
\citealt{arcavi2010,li2011b,smith2011,eldridge2013}),
although Type Ibc SNe are slightly overproduced
(see also discussion on uncertainties in Section~\ref{unce}).
When the dividing mass between Type II and Type IIb is set to 2, 0.5,
and 0.1~\Msun,
the SN~IIb fractions become 2.9\%, 2.2\%, and 2.0\%, respectively
(see also \citealt{claeys2011}).
The fraction of hydrogen-free SNe is not affected by the dividing
mass of the two hydrogen-rich SN types.

To summarize, our population synthesis model reproduces bulk observational
properties of core-collapse SNe, namely, the rate and the fraction of the
hydrogen-rich and hydrogen-free SNe, reasonably well. Thus, the model is presumed
to represent the binary properties of SN progenitors reasonably.

\subsection{Uncertainties}\label{unce}
Because the number of stars used in our population synthesis calculations
is sufficiently large, the errors in the estimated fractions below
are mainly systematic.
To evaluate the systematic uncertainties, we perform the binary
population synthesis calculations with several model assumptions.
The largest deviation in the predicted fractions we found
is by a factor of 1.5 from the standard fraction shown in the next section
when we reduce the common-envelope ejection efficiency $\alpha_\mathrm{CE}$
to 0.5 from the standard value $\alpha_\mathrm{CE}=1$.
We also refer to \citet{demink2014,claeys2011,claeys2014} for the recent
investigations on the systematic uncertainties in the binary evolution
modelling with similar algorithm.

\begin{table}
\caption{Fractions predicted from our population synthesis model. See Section
 \ref{unce} for uncertainties.} 
\label{table1}      
\begin{center}         
\begin{tabular}{c c c c}
\hline       
SN type & SN fraction  &$f_\mathrm{e}$(binary)$^a$ & $f$(observable)$^b$ \\
\hline                    
 II  & 0.54 & 0.29 & $3.8\times 10^{-5}$\\
 IIb & 0.10 & 0.26 & $1.6\times 10^{-4}$ \\
 Ibc & 0.36 & 0.61 & $5.3\times 10^{-3}$ \\
\hline
 all types & 1 & 0.40 & $1.9\times 10^{-3}$\\
\hline
\end{tabular}
\end{center}
{\footnotesize
$^a$ Fraction of SNe in binary systems at the time of the
 explosion in the given SN type. 
\\
$^b$ Fraction of SNe in which the collisional early LC
 brightening is observable in the given SN type.
}
\end{table}

\section{Predicted observable fractions of the early light-curve brightening due to the collision}\label{fraction}
Based on our population synthesis model, we estimate
the expected fractions of core-collapse SNe in which
the LC brightening due to the binary collision is observable.
Our population synthesis model provides
the fractions $f_s$ of SNe coming from each simulated progenitor system
at the time of the explosion.
The total fraction of SNe in binary systems when they explode is expressed as
\begin{equation}
 f_\mathrm{e}(\mathrm{binary})={\displaystyle \sum^{}_{\mathrm{binary}}} f_s.
\label{fracsum}
\end{equation}
The summation in Eq.~(\ref{fracsum}) is taken for all the SN progenitors
which are in binary systems at the time of the explosion.
The binary fractions predicted from our population synthesis model are shown in
Table~\ref{table1}.
The binary fractions are smaller at the time of the explosion
than at the time of the birth
because a SN progenitor can leave the binary system
when its companion star explodes earlier.
Stellar mergers also reduce the binary fractions.
Because SN~II progenitors tend to have longer life times, their binary fraction
is more reduced when they explode than that of SN~Ibc progenitors.
The binary fraction is larger in SNe~Ibc because
more mass can be lost in the binary systems because of
the binary interactions during the evolution,
allowing the progenitors in the binary systems
to strip the hydrogen-rich envelope more easily.
The binary evolution makes the predicted
fraction of the stripped SNe close to that observed
(e.g. \citealt{izzard2004,eldridge2008,eldridge2011}).

We assume that LC brightening from a binary collision
can be observed if the observer is within the half opening angle
$\theta_h$ of the hole created by the collision (K10).
Following K10, we set
\begin{equation}
 \theta_h\simeq \arctan \left(\frac{2R_\star}{a}\right),
\end{equation}
where $R_\star$ is the radius of the companion and $a$ is the separation.
Fig.~\ref{collision} presents a schematic picture of these parameters.
If the observer is at a random location, the probability $p_h$
that the SN is observed within the half opening angle is
\begin{equation}
p_h=\frac{\Omega_h}{4\pi} =\frac{1-\cos\theta_h}{2},\label{eqprob}
\end{equation}
where $\Omega_h$ is the solid angle of the hole.
The fraction of SNe in which the collisional brightening can be observed is
\begin{equation}
 f(\mathrm{observable})={\displaystyle \sum^{}_{\mathrm{binary}}} f_s p_h.
\label{obsfraction}
\end{equation}

\begin{figure}
\centering
\includegraphics[width=\columnwidth]{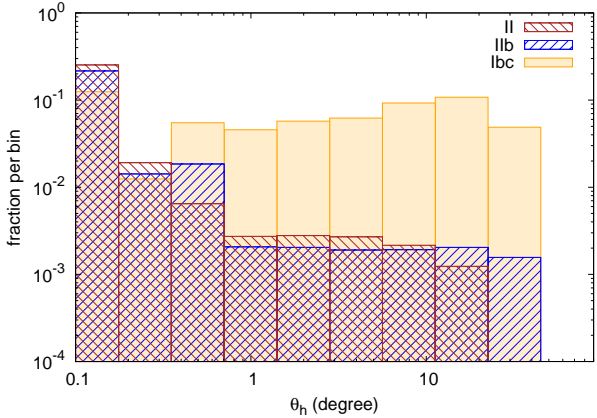}
\caption{
Hole half-opening angle ($\theta_h$) distribution predicted from our
population synthesis model.
The fractions of the SNe in the given
 SN type in each angle bin are presented. The leftmost bin includes
all the SNe below $\theta_h=0.18^\circ$.
}
\label{angledistribution}
\end{figure}

The fractions of SNe in which the LC brightening from a collision
is observable are summarized in Table~\ref{table1}:
$0.19$\% of all core-collapse SN LCs are found to be brightened by the collision.
The observable fractions in SNe~II and IIb are small.
The expected fraction for SNe~II is $3.8\times 10^{-3}$\%.
The fraction is 
larger in SNe~IIb ($1.6\times10^{-2}$\% for the standard model).
The dividing mass between SNe~II and IIb is found to have a small
effect on the observable fraction
($1.8\times10^{-2}$\% for 2 \Msun, $1.5\times10^{-2}$\% for 0.5 \Msun, and $1.4\times10^{-2}$\% for 0.1 \Msun).
Fig.~\ref{angledistribution} shows the distributions of $\theta_h$
for each SN type in our standard model.
Although the binary fractions are similar
in SNe~II and IIb in our population synthesis model, SN~IIb progenitors
have larger $\theta_h$, making the observable fraction larger.
This is because hydrogen-rich SN progenitors in closer binary
systems can lose their envelope more easily by the binary interaction during the evolution.
Compared to the hydrogen-rich SNe, SNe~Ibc have a much larger observable fraction, 
0.53\%, thanks to the larger binary fraction and the flatter
distribution of $\theta_h$ (Fig.~\ref{angledistribution}).
In other words, we expect to observe the brightening in about $5$
out of 1000 SNe~Ibc with early discoveries.

\section{Discussion}\label{discussion}
\begin{figure}
\centering
\includegraphics[width=\columnwidth]{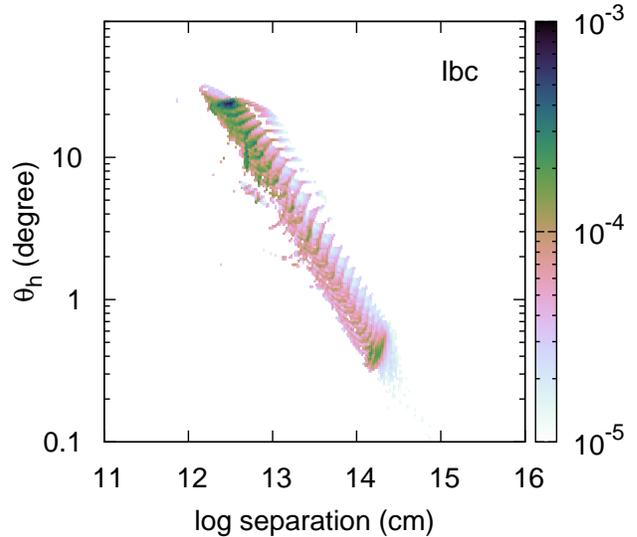}
\caption{
Fractions of SNe~Ibc in the separation-$\theta_h$ plane
at the time of the explosion.
The fraction of SNe~Ibc in each bin is indicated by colour.
Nothing is plotted in the regions with a fraction smaller than $10^{-5}$.
}
\label{sepangleibc}
\end{figure}

We have shown that the fractions of core-collapse SNe in which the 
LC brightening from the binary collision can be observed are very small.
The predicted small observational probabilities disfavour
the argument that the early slow LC decline observed in SN~Ib~2008D is
because of the collision with the binary \citep{dessart2011},
given the total number of the well-observed SN~Ibc LCs is about 100
(e.g. \citealt{li2011b,drout2011,bianco2014,lyman2014,taddia2014}).
\citet{bersten2013} argued that the separation needs to be
$a\simeq 10^{11}$~cm to explain the early LC of SN~2008D by the
collision based on their estimates of the SN properties.
Our population synthesis result indicates that 
the systems with $a\simeq 10^{11}$~cm in SNe Ibc are extremely rare
(Fig.~\ref{sepangleibc}), because this separation is the typical radius
of Wolf-Rayet stars.
This further suggests that the early slow LC
decline is not likely because of the collision.
Other mechanisms are likely to
be related to the mysterious slow LC decline of SN~2008D in its early epoch.

Except for SN~2008D, there are no reported SNe~Ibc with
the clear early slow LC decline (e.g. \citealt{li2011b,drout2011,bianco2014,lyman2014,taddia2014}).
This is consistent with the estimates from our population synthesis.
SN~Ibc LCs with the observations early enough
to see the collisional effect are still rare.
Our prediction that 0.53\% of SNe~Ibc
are brightened by the collision can be tested by future large SN surveys.
For example, the Large Synoptic Survey Telescope expects $\sim 10^4$
SN~Ibc discoveries per year \citep{lsst2009} and it will find
$\sim 10$ SNe~Ibc with the early brightening by the collision
per year if they are detected early enough.
The Zwicky Transient Facility is also likely to be capable of detecting
such event thanks to their emphasis on the early SN discovery \citep{bellm2014}.

On the contrary, the early slow LC declines of SNe~IIb are frequently
observed (e.g. \citealt{bufano2014})
and the observed SN fraction with the early slow LC decline is presumed to be much
larger than the fraction estimated here.
\citet{dessart2011} argued that the extremely large radii required to explain
the early slow LC decline observed in SNe~IIb
(e.g. $\gtrsim 630\ \Rsun$ for SN~1993J, \citealt{blinnikov1998})
may indicate that they are actually brightened by the binary collision.
However, the predicted fractions in our model
to observe the collisional brightening
in SNe~IIb ($\sim 0.01$\%) are even smaller
than those in SNe~Ibc ($\sim 0.1$\%).
Thus, the slow LC declines often observed in SNe~IIb are presumed 
to be mostly from another cause, e.g. the extended
early cooling phase caused by the large progenitor radii.

It is interesting to note that
SNe~IIb are suggested to come from both compact and extended
progenitors \citep{chevalier2010}. 
If we can identify the compactness of SN~IIb progenitors by
multi-wavelength observations
like the radio observations demonstrated by \citet{chevalier2010}
and the early slow declines
are associated with the compact progenitors, 
the early slow declines are more likely to be related to the collision
because of the rapid adiabatic cooling of the compact SN ejecta.
In our standard population synthesis model, about $10$\% 
of SNe~IIb have progenitors with radii smaller than 10~\Rsun.
This fraction increases as the dividing mass decreases because
closer binary systems tend to lose more envelope mass.
When the dividing mass is 0.1~\Msun, about 60\% of the SN IIb
progenitors have radii shorter than 10~\Rsun.
This kind of information in SNe~IIb may be used to understand the
physics involved in the binary evolution (cf. \citealt{claeys2011,stancliffe2009}).

An interesting feature of the LCs affected by the collision is the
strong dependence of the luminosity on the separation (K10).
The early isotropic bolometric luminosity $L_{\mathrm{c,iso}}$
by the collision observed within $\theta_h$
is formulated by K10 as,
\begin{eqnarray}
L_{\mathrm{c,iso}}&\simeq& 2\times 10^{43}\
\left(\frac{a}{10^{13}\ \mathrm{cm}}\right)
\left(\frac{M_\ej}{10\ \Msun}\right)^{\frac{1}{4}} 
\left(\frac{v_\ej}{10^9\ \mathrm{cm~s^{-1}}}\right)^{\frac{7}{4}}
\nonumber \\
&&
\ \ \ \ 
\times\left(\frac{\kappa_e}{0.2~\mathrm{cm^2~g^{-1}}}\right)^{-\frac{3}{4}}
\left(\frac{t}{\mathrm{day}}\right)^{-\frac{1}{2}}
\ \ \mathrm{erg~s^{-1}},
\end{eqnarray}
where 
$M_\ej$ is the SN ejecta mass,
$v_{\ej}$ is the
SN ejecta velocity colliding to the binary star,
$\kappa_e$ is the electron
scattering opacity in the SN ejecta, and $t$ is the time
since the explosion.
The luminosity is proportional to the separation because 
the collisional heating at longer separations is less affected by the subsequent
adiabatic cooling (K10).
In Fig.~\ref{sepangleibc}, we plot the fraction distribution of the
SN~Ibc progenitors in binary systems at the time of the explosion.
The angle $\theta_h$ is plotted vs the separation.
The angle indicates the probability for the system to be observed
within the hole created by the collision (Eq. \ref{eqprob}).
The most probable systems to be
observed have $a_{13}= 0.1-1$. This indicates that the early luminosity 
affected by the collision is most likely to be
$\sim 10^{42-43}$ $\mathrm{erg~s^{-1}}$ for SNe~Ibc
with a typical ejecta mass of 2 \Msun\ \citep{drout2011}.
Because the binary configuration of the most probable systems is similar to
the system with which the LC brightening is simulated by K10
in terms of the ratio of the separation to the companion radius
(a factor of a few),
the SN~Ibc LCs with the collisional brightning are presumed to be similar to those in K10.
For SNe~II and IIb, the separations of the systems with the highest observable
probabilities are $\sim 10^{14}$ cm. Thus, the typical early time
luminosities from the collision can be $\sim 10^{44}$
$\mathrm{erg~s^{-1}}$ with $M_{\ej}=10\ \Msun$,
although this is rarely observed.

\section{Conclusions}\label{conclusion}
We investigate the expected fractions of core-collapse SNe in
which the LC brightening due to the collision between the SN ejecta and the
companion star is observable by
performing population synthesis simulations
with binary stellar evolution.
The brightening of the LCs is assumed to be observable if an
observer is within the direction of the hole created by the collision
(K10, Fig.~\ref{collision}).

The expected fractions are summarized in Table~\ref{table1}.
We find that only 0.53\% of SN~Ibc LCs can be observed with the brightening
because of the binary collision, although 61\% of them are in binary systems when
they explode.
The observable fractions in hydrogen-rich SN LCs are much smaller because of
the smaller fractions of the progenitors in binary systems at the time
of the explosions and
the larger separations ($\sim10^{-3}$\% in SNe~II and $\sim10^{-2}$\% in SNe~IIb).

The very small predicted fractions disfavour
the interpretation that the early slow LC decline
observed in SN~Ib~2008D is due to the binary collision.
The separation required to explain the early LC is also unrealistically
small (Fig.~\ref{sepangleibc}, see also \citealt{bersten2013}).
We find that the most probable systems to be observed with LC brightening
have $a=10^{12}-10^{13}$~cm (SNe~Ibc) and $a\sim10^{14}$~cm (SNe~II
and IIb). Thus, the expected luminosities of the LCs affected by the collision
are $\sim10^{42}-10^{43}$ $\mathrm{erg~s^{-1}}$ (SNe~Ibc) and
$\sim10^{44}$ $\mathrm{erg~s^{-1}}$ (SNe~II and IIb).

The small expected fractions indicate that
the slow LC declines observed so far are likely to be caused
by other mechanisms than the collision.
However, future large transient surveys such as
the Large Synoptic Survey Telescope
or the Zwicky Transient Facility
will be able to observe SN LCs with the
collisional brightening.
The Large Synoptic Survey Telescope is expected to detect $\sim10$ SNe~Ibc
with the early brightening due to the collision per year.
The Zwicky Transient Facility is also capable of detecting
such events with their emphasis on early SN discovery.

\section*{Acknowledgments}
We thank the referee for the comments which improved this paper.
TJM is supported by Japan Society for the Promotion of
 Science Postdoctoral Fellowships for Research Abroad
 (26\textperiodcentered 51).
ZWL is supported by the Alexander von Humboldt Foundation.
RGI thanks the Alexander von Humboldt Foundation for funding his
 position, John Eldridge for useful discussion, and
the STFC for his Rutherford Fellowship.

\label{lastpage}

\end{document}